# Edward Schenfeld: Visual Photometry of Variables: I. Comparison Stars

*V. I. Burnashev, B. A. Burnasheva*

Crimean Astrophysical Observatory, pos. Nauchnyi, 98409 Crimea, Ukraine

**Abstract.**
Based on the SIMBAD database, we collected necessary information on the comparison stars used by E. Schenfeld in observations of variable stars in the 19th century.

Full privy councilor, Professor of Bonn University Edward Schenfeld (1828-1891) in collaboration with F.W.A. Argelander, compiled the catalog *Bonner Durchmusterung* (Bonn Survey, abbreviated BD). He was recognized by the astronomer community as a highly qualified and careful observer. Owing to Dr. Schenfeld's efforts, the interests of His Majesty the Duke, the government, and the university and municipal administrations united to establish Heidelberg Observatory in Bonn. The untimely decease of E. Schenfeld in 1891 impeded publication of his observations.

The results of his research became familiar to the general public owing to the efforts of W. Valentinier, who have taken the place of Edward Schenfeld as the director of Heidelberg Observatory. Estimates of the magnitudes of various variable stars carried out by Schenfeld at Mannheim and Karlsruhe in 1865-1875 were published in the first volume of *Veroff. Sternwarte Heidelberg* (Valentinier, 1900).

E. Schenfeld's observations before 1865 were mainly focused on nebulas. In particular, he compiled a catalog of 489 nebulas. At the same time, he occasionally estimated the magnitudes of certain variables, e.g., S Cnc, R Mon, β Per, ρ Per, and R Per since 1861; R Psc since 1863; and S Vir, T Vir, and V Vir since 1860.

His observations included comparison of the magnitudes of 118 variables with about 1100 other stars. According to Schenfeld's estimates, 35963 complete observations were carried out in 1865-1875, including 80000 individual estimates, apart from individual comparison of stars (about 5000). Some stars he observed after moving to Bonn, e.g., o Cet (until 1889), β and ρ Per (until 1888), and δ Cep (until 1888).

The first Chandler catalog of variable stars did not include R Cep, δ Ori, and V Sgr; however, according to Schenfeld, their brightness's vary by up to half a magnitude ($0^m.5$-$0^m.6$). Hence, variability of these stars was discovered for the first time by Schenfeld.

For the user's convenience, Schenfeld's observations presented in *Veroff. Sternwarte Heidelberg* were divided into two parts. The first part comprises estimates of variable-star magnitudes with respect to comparison stars by the Argelander stepwise estimation method. The second part gives information on comparison stars.

All necessary reductions are performed only for two stars. Observations of β Per in 1869-1875 were reduced by J. Scheiner, and the data for the other variable, η Aql, by W. Lockyer. If not specifically noted otherwise, the star coordinates are given for the epoch 1855.0 to facilitate identification using the BD.

Table 1 gives the list of variable stars observed by Schenfeld. The type of variability adopted in the 4th edition of *The General Catalog of Variable Stars* (GCVS). The list is overwhelmed by pulsating stars, including 82 Mira variables (M according to the GCVS classification), 6 cepheids (DCEP), 1 RR Lyr star (RRAB), 7 red semiregular variables (SR), 2 irregular variables (LB), and 2 RV Tau stars (RV). The list includes eruptive variables: one RCB (R CrB) star, one novalike star (T CrB), one UG (U Gem) star, and three Orion variables. Two stars did not exhibit brightness variations (CST) and another one (o Tau) is most probably not a variable either.

**Table 1.** List of variable stars observed by E. Schenfeld in the 19th century

| | | | | | | | |
|---|---|---|---|---|---|---|---|
| R Andromedae | M | δ Cephei | DCEP | S Hydrae | M | R Sagittarii | M |
| R Aquarii | M | R | CST | T | M | S | M |
| S | M | o Ceti | M | R Leonis | M | T | M |
| T | M | R | M | S | M | U | DCEP |
| η Aquilae | DCEP | S | M | R Leonis Min. | M | V | CST |
| R | M | S Comae Ber. | RRAB | R Leporis | M | R Scorpii | M |
| S | SRA | R Coronae Bor. | RCB | δ Librae | EA/sd | S | M |
| T | LB: | S | M | R | M | R Scuti | RVA |
| R Arietis | M | T | NR | S | M | R Serpentis | M |
| S | M | U | EA/sd | R Lyncis | M | S | M |
| T | SRA | R Corvi | M | β Lyrae | EB | T | M |
| ε Aurigae | EA/gs | χ Cygni | M | R Monocerotis | INA | λ Tauri | EA/dm |
| R | M | R | M | S | IA | o | ? |
| R Bootis | M | S | M | T | DCEP | R | M |
| S | M | T | LB | R Ophiuchi | M | S | M |
| R Camelopardalis M | M | U | M | S | M | T | NT |
| R Cancri | M | R Delphini | M | δ Orionis | EA/dm | V | M |
| S | EA/ds | S | M | R | M | R Ursa Majoris | M |
| T | SRB | T | M | S | M | S | M |
| U | M | ζ Geminorum | DCEP | R Pegasi | M | T | M |
| V | M | η | SRA + EA | S | M | R Virginis | M |
| R Canis Minoris | M | R | M | T | M | S | M |
| S | M | S | M | β Persei | EA/sd | T | M |
| T | M | T | M | ρ | SRB | U | M |
| R Capricorni | M | U | UGSS + E | R | M | V | M |
| T | M | R Herculis | M | S | SRC | W | CWA |
| U | M | S | M | R Piscium | M | R Vulpeculae | M |
| R Casssiopejae | M | T | M | S | M | S | DCEP |
| S | M | U | M | T | SRB | | |
| T | M | R Hydrae | M | R Sagittae | RVB | | |

At present, accurate photoelectric measurements exist for many comparison stars. Thus, we attempted to combine all Schenfeld observations in a unified system and made it accessible to the general research community. The first part of this work includes identification of comparison stars and the collection of relev ant data. Dr. Jost attempted to identify the comparison stars based on records, sketches, etc., in observation logs. However, the coordinates of some comparison stars could not be found. The average accuracy of BD-star coordinates given by Valentinier is about 2-3 arcmin. The accuracy for faint stars is worse, up to 5 arcmin.

Relatively bright stars up to about 10th magnitude were identified and their numbers in the BD, HD, SAO, Tycho, and Hipparcos catalogs were determined using the SIMBAD database (Set of Identifications, Measurements and Bibliography for Astronomical Data, SIMBAD Query Form -

http://www.simbad.u-strasbg.fr/simbad/sim-fid) and the Astronet database developed at SAI (http://www.astronet.ru/db/map/ ).

The database VisieR Service (http://www.vizier.u-strasbg.fr/) included in the SIMBAD database provides for information on fainter-star magnitudes. Since the VisieR Service employs many catalogs, there exists the possibility of independently estimating the reliability of stellar magnitudes in various catalogs.

Even a brief comparison of photometric evaluations by different authors shows that the magnitudes in different catalogs, even the visual magnitudes, can differ by up to half a magnitude. Of course, these differences can to a certain extent be explained by variability. Nevertheless, even visual magnitudes of a number of common stars from different catalogs differ on average by $0^m.1$-$0^m.2$.

Note that in contrast to brighter stars whose SIMBAD magnitudes are certain weighted averages, the astrometric and photometric data for faint stars were taken from the NOMAD (Naval Observatory Merged Data Set) database.

Therefore, for definiteness, only the NOMAD magnitudes are given for faint stars (Zacharias et al., 2004). In our opinion, this data set is the most comprehensive available (up to the mid-2007). In cases where stars could not be found in NOMAD, the data from the All-Star Compiled Catalog (ASCC) were used (Kharchenko, 2001). The average mismatch between the former and latter catalogs (Kharchenko, 2001) is about $0^m.01$, i.e., much smaller than the error in determining the magnitude of a star from the Schenfeld list. The visual magnitudes of a few stars were taken from the Guide Star Space Telescope (GST) (http://www-gss.stsci.edu/) or The Amateur Sky Survey (TASS) (Droege et al., 2006) catalogs if the mentioned catalogs were missing the corresponding information. In a few cases where information on the visual magnitude of a star was absent in the above-mentioned catalogs, the $V$ magnitudes were estimated using the USNO-A data (http://www.iac.es/galeria/mrk/comets/USNO/Landolt.htm).

It should also be noted that in some cases, failure to identify all the stars can probably be explained by misprints or erroneous identification by the author. Using the figures in the text, we identified certain comparison stars, e.g., $d = c$ and $b$ for T Cas.

**Table 2.** 1. AND R

| 1 | 2 | 3 | 4 | 5 | 6 | 7 | 8 | 9 | 10 | 11 | 12 | 13 |
|---|---|---|---|---|---|---|---|---|---|---|---|---|
|   | BD | HD | B | V | sp | PPM | SAO | TYC | HIP | α(2000) δ(2000) | m(B) m(R) | Remark |
| $R$ | +37° 58 | 1967 | $9^m.36$ | $7^m.39$ | Se | - | 53860 | - | 1901 | | | |
| $p$ | +40° 29 | 905 | $6^m.03$ | $5^m.72$ | F0IV | 42707 | 36173 | - | 1086 | | | |
| $n$ | +39° 56 | 1527 | $7^m.538$ | $6^m.359$ | K1 III | 42818 | 36269 | 2786-272-1 | 1575 | | | |
| $l$ | +37° 48 | 1712 | $7^m.959$ | $6^m.866$ | K0 | 65229 | 53835 | 2783-927-1 | 1715 | | | |
| $k$ | +36° 39 | 1770 | $8^m.47$ | $7^m.51$ | G5 | 65242 | 53842 | 2783-1711-1 | 1760 | | | |
| $c$ | +37° 53 | - | $10^m.957$ | $9^m.996$ | - | - | - | 2783-1407-1 | - | $00^h22^m25^s96 + 38°23'27''5$ | | |
| $m$ | +37° 54 | 1831 | $8^m.303$ | $6^m.745$ | M0 | 65251 | 53849 | 2783-1160-1 | 1809 | | | var |
| $b$ | +37° 55 | - | $11^m.039$ | $10^m.929$ | - | - | - | 2783-900-1 | - | $00^h23^m10^s33 + 38°26'58''4$ | | |
| $d$ | +37° 57 | 1905 | $10^m.142$ | $10^m.025$ | A3 | 65259 | - | 2783-576-1 | - | $00^h23^m32^s52 + 38°34'49''7$ | | |
| $q$ | +37° 57B | - | $12^m.739$ | $12^m.003$ | - | - | - | - | - | $00^h23^m32^s78 + 38°35'52''7$ | | double |
| $g'$ | +36° 45 | 1906 | $8^m.49$ | $8^m.04$ | F8 | 65260 | 53856 | 2273-1359-1 | 1861 | | | |
| $y$ | - | - | $12^m.551$ | $11^m.570$ | - | - | - | - | - | $00^h24^m02^s31 + 38°31'33''7$ | | |
| $x$ | - | - | $12^m.904$ | $12^m.305$ | - | - | - | - | - | $00^h24^m06^s28 + 38°33'30''8$ | | |
| $z$ | - | - | $13^m.210$ | $12^m.330$ | - | - | - | - | - | $00^h24^m17^s01 + 38°30'09''7$ | $13^m.4b$ $12^m.2r$ | |
| $e$ | +37° 59 | - | - | $9^m.484$ | - | - | - | - | - | | | |
| $g$ | +38° 46 | 2153 | $8^m.99$ | $8^m.67$ | F0 | 65294 | 53885 | 2783-1992-1 | 2033 | | | double |
| $f$ | +37° 65 | 2209 | $9^m.89$ | $9^m.21$ | A2 | 65306 | 53892 | 2783-1698-1 | - | | | |
| $h$ | +37° 68 | 2312 | $9^m.20$ | $7^m.67$ | K0 | 65326 | 53904 | 2783-925-1 | 2145 | | | |
| $p'$ | +38° 90 | 3817 | $6^m.22$ | $5^m.33$ | G8III | 65538 | 54079 | - | 3231 | | | |

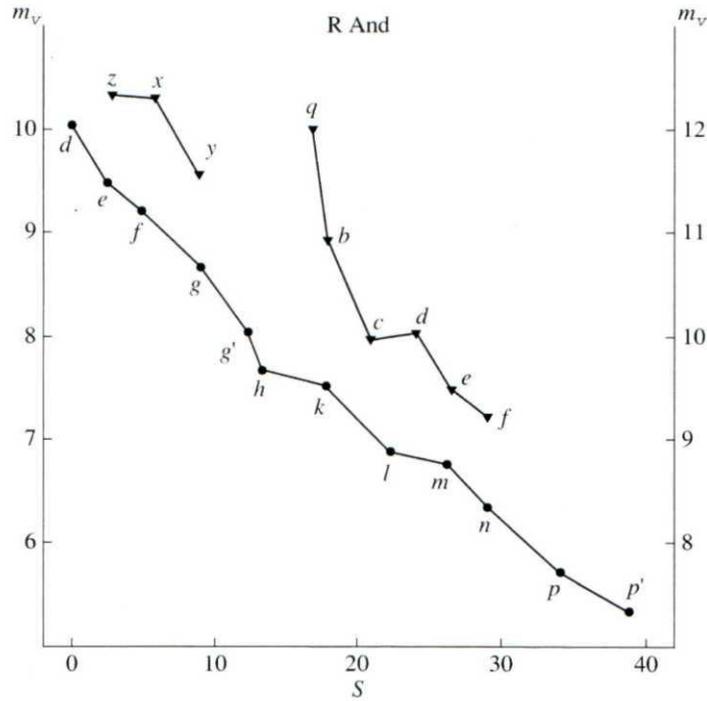

**Fig 1.** The Power-law scale plotted using the comparison stars from the Schenfeld list for R And. The left- and right-hand magnitude scales correspond to the lower and upper curves, respectively.

The data are collected in Table 2. As an example, we present a record from the table for the first variable, R And. The corresponding plot reducing the Schenfeld power-law scale to *V* magnitudes is shown in Fig 1. The left-hand (brighter star) and right-hand magnitude scales correspond to the lower and upper curves, respectively.

The first column of the table is the designation introduced by Schenfeld for each star; the 2nd and 3rd columns are the star numbers according to the BD and HD catalogs, respectively. The photoelectric *B* and *V* magnitudes are listed in the 4th and 5th columns, while the 6th column gives the spectral class. The PPM, SAO, TYC, and HIP catalog numbers of the stars can be found in the 7th, 8th, 9th, and 10th columns, respectively.

In those, unfortunately more than a few, cases where the star is not found in any of the above-mentioned catalogs, its coordinates (epoch 2000) are placed in the 11th and 12th columns, respectively. The 12th column gives the magnitudes of some stars estimated using the maps of the Palomar all-sky survey in the photographic (*b*) and red (*r*) bands. Finally, the 13th column is allocated for additional information such as variability, binary nature, etc.

## NOTES TO TABLE 2

In addition to notes in the 13th column of Table 2, additional information on some stars, numbered according to the Schenfeld list, are given below. If not explicitly stated otherwise, the notes on the coordinates concern the coordinates published by Schenfeld for equinox 1855.0.
1. R And:
   − *f* : a declination error of 5': +37°32:0 should be replaced by +37°37'0;
   − *x, y,* and *z*: deviations of up to 3'.

2. R Aqr:
   – *a* is a binary. The magnitudes of both components are given. The total magnitudes are $B = 10^m.75$ and $V = 10^m.53$;
   – there may be confusion with the nomenclature: *h'* = *g* and *c'* = *b*.
3. S Aqr:
   – *d:* the right ascension may be erroneous: $22^h51^m39^s.8$ should be replaced by $22^h51^m09^s.8$.
4. T Aqr:
   – *p:* the coordinates are absent;
   – *k:* coordinate or magnitude error.
6. R Aql:
   – *p:* misprinted BD number: BD + 8 3969 should be replaced by BD + 8 3968;
   – *s*: not identified. It seems that neither *s* nor *s'* are suitable in terms of both magnitude and coordinates.
7. S Aql:
   – *b:* a declination error of 5': +15°16'.6 should be replaced by +15°11'.6;
   – *d:* not suitable in terms of magnitude;
   – *x* and *y*: not identified.

The last three stars in the list are designated only by the BD numbers: BD + 15 4082 = RW Aql, BD + 15 4074 A and BD + 15 4080.

11. T Ari:
    – *m*: misprinted declination: +15°23'8 should be replaced by: +16°23'8;
    – *d* = AM Ari.
13. R Aur:
    – *d:* erroneous declination: +53°29'.6 should be replaced by +53°19'6;
    – *s:* not identified. Possibly erroneous identification of the stars *r* and *t*. It is possible that *r* = *r'*.
14. R Boo:
    – *a* : the coordinates are absent in the list of comparison stars.
    – *m:* a declination error of 10': +27°27'.5 should be replaced by +27°37'.5.
15. S Boo:
    – *l'* and *k* are variables.
20. U Cnc:
    – *h*: erroneous declination: +19°20'.3 should be replaced by +19°15'.3.
21. V Cnc:
    – *h*: erroneous BD number: BD + 17 1726 should be replaced by BD + 17 1826;
    – *g*: a declination error of 10 arcmin: +18°01'4 should be replaced by +17°51'.4.
22. R CMi:
    – *o*: not designated in the list of comparison stars on page 263;
    – *c*: erroneous BD number: BD + 10 1430 should be replaced by BD + 10 1432.
23. S CMi:
    – misprinted nomenclature on page 263: η should be replaced by *c*.
25. R Cap:
    – probable confusion of the names: *d* is written instead of *n'* and vice versa;
    – *n'* = BD - 14 5646: does not fit in terms of magnitude;
    – *d*: also does not fit in terms of magnitude, possibly a variable;
    – *e*: the magnitude V = 9.742 was taken from the ASCC catalog;
    – *b*: the magnitude V = 12.0 was estimated using the USNO data.

27. U Cap:
  – coordinates of the stars a, b, f, g, and h are not given;
  – *e*: V = $10^m.9$ according to ASCC data.
29. S Cas:
  – *a*: a binary with the total magnitudes B = $7^m.61$ and V = $6^m.9l$;
  – *n* = V762 Cas.
30. T Cas:
  – *d* = *c* and *b* are identified using sketches in the text;
  – *l*: erroneous declination: +58°24'.8 should be replaced by +58°34'.8.
32. R Cep = UZ UMi.
34. R Cet:
  – *c*, *d*, and *a*: errors of up to 4'-5'.
35. S Cet:
  – Nova: possible coordinates are α(2000)= $00^h24^m13^s$ ± $3^s.8$ (Δα(max) ≈ $1^s.3$), δ(2000) = -08°56'.6 ± 10' (Δδ(max) ≈ 21').
37. R CrB:
  – *k*: the TASS magnitude V = $12^m.321$.
38. S CrB:
  – *k'* = U CrB before 1867.
40. U CrB:
  – the comparison stars are the same as for S CrB.
41. R Crv:
  – *a, c, e,* and *g* are far away, an error of 4'-6' is possible;
  – *a*: the magnitude V = $11^m.7$ according to USNO data.
42. χ Cyg:
  – *t*: erroneous declination: +32°28'.1 should be replaced by + 32°38'.1;
  – *u*: a binary, the USNO catalog gives the magnitude V = $11^m.5$ for *u*(A) and the total magnitudes *u*(A + B): B = $11^m.2$ and V = $11^m.25$;
  – *p'*: erroneous BD number: BD + 32 3582 should be replaced by BD + 32 3583;
  – *v*: one of the following candidate stars:
     (1) PPM 83664 with α(2000) = $19^h50^m06^s.38$, δ(2000) = +32°52'51".00, B = $12^m.108$, and V = $11^m.549$ (The data for this star are given in Table 2);
     (2) PPM 83663 with α(2000) = $19^h50^m35^s.29$, δ(2000) = +32°50'46".09, B = $12^m.247$, and V = $11^m.988$.
43. R Cyg:
  – *q*: erroneous BD number: BD + 49 3071 should be replaced by BD + 49 3072.
45. *T* Cyg:
  – *l*: erroneous BD number: BD + 34 4069.
46. U Cyg:
  – *g*: a triple star.
47. R Del:
  – *l*: possibly misidentified or a large-amplitude variable.
48. S Del:
  – *g*: a visual binary:
     (1) α(2000) = $20^h43^m27^s.90$, δ(2000) = +16°56'42".01, B = $9^m.589$, and V = $9^m.139$.
     (2) α(2000) = $20^h43^m27^s.91$, δ(2000) = +16°56'38".01, B = $9^m.993$, and V = $9^m.503$.
     The total magnitude V = $8^m.48$.

52. R Gem:
   – $w^2$: erroneous BD number: BD + 22 1556 should be replaced by BD + 22 1566;
   – $a$: erroneous BD number: BD + 22 1596 should be replaced by BD + 22 1569.
53. S Gem:
   – $y, l,$ and $x$: unreliably. Large coordinate errors up to 5'-7' are possible.
54. T Gem:
   – $k$: erroneous BD number: BD + 24 1767 should be replaced by BD + 24 1766.
55. U Gem:
   – $n$: right ascension can be erroneous: $07^h48^m13^s$ should be replaced by $07^h48^m43^s$.
56. R Her:
   – $a$: a binary with the total magnitudes $B = 11^m.42$ and $V = 11^m.10$;
   – $q$: the magnitude $V = 12^m.53$ was taken from the GSC catalog.
58. T Her:
   – $e$: erroneous declination: +30°39'.9 should be replaced by +30°40'.9;
   – $l$: erroneous BD number: BD + 31 3216 should be replaced by BD + 31 3296.
59. U Her:
   – $l$: erroneous BD number: BD + 16 2493 should be replaced by: BD + 16 2943;
   – $c$: a declination error of 20 arcmin: +19°16'.8 should be replaced by +19°36'.8;
   – $a$: the magnitude $V = 12^m.35$ was taken from the GCS catalog.
60. R Hya:
   – $a$ = CD - 24 10174 = CPD - 24 4786.
61. S Hya:
   – $f$: designated twice in the text by the same letter corresponding to different coordinates;
   – $x = p$. $V = 12^m.08$ was taken from the GCS catalog;
   – $a$ and $b$: probably confused with each other in the text since $b$ should be brighter than $a$.
62. T Hya:
   – $i$: a binary:
      (1) $\alpha(2000) = 08^h55^m33^s.75$, $\delta(2000) = -09°03'10''.08$, $B = 12^m.960$, and $V = 12^m.840$.
      (2) $\alpha(2000) = 08^h55^m35^s.55$, $\delta(2000) = -09°03'19''.09$, $B = 13^m.700$, and $V = 13^m.580$.
      The total magnitudes are $B = 12^m.51$ and $V = 12^m.40$.
71. $bet$ Lyr = SAO 67451 + SAO 67452.
76. S Oph:
   – $h$: DOES NOT FIT AT ALL! Probably a large-amplitude variable;
   – $a'$ and $k$: present in the text, but absent in the list of comparison stars; identified by the author.
79. S Ori:
   – $d$: erroneous BD number: BD - 05 1258 should be replaced by BD - 05 1268;
   – $m$ and $k$: no information available for these stars in the text of the table and in page 268;
   – $k$: possible identification using brightness of nearby stars;
   – $o, p,$ and $n$: absent at all, identified using the sketches in the text.
80. R Peg:
   – $\zeta$: erroneous BD number: BD + 10 4994 should be replaced by BD + 10 4894;
   – $\alpha'$: erroneous BD number BD: BD + 8 4968 should be replaced by BD + 10 4873AB;
   – $b$: a binary.
81. S Peg:
   – $p$: erroneous declination: +8°51' should be replaced by +8°41';
   – $o$: large disagreement between the data of various catalogs: $V = 15^m.030$ (NOMAD), $V = 12^m.93$ (GSC).
82. T Peg:
   – probable confusion of the stars $a$ and $b$ in December 1872.

85. R Per:
   – *a:* the magnitude $V - 12^m.08$ was taken from the GCS catalog.
86. S Per:
   – *h:* not designated, probably confused with *k*;
   – *m, l,* and *k:* not designated, identified using sketches in the text;
   – *n:* not identified;
   – *g:* erroneous BD number: BD + 57 596. The star is probably not designated in this catalog;
   – *c:* a binary with the total magnitudes $B = 7^m.92$ and $V = 7^m.70$.
87. R Psc:
   – *n, o,* and *p:* no information available for these stars which remained unidentified;
   – *c':* right ascension can be erroneous: $01^h18^m02^s.3$ should be replaced by $01^h18^m12^s.3$.
88. S Psc:
   – *z:* erroneous BD number: BD + 7 194. This star is probably not pointed out in this catalog.
90. R Sge:
   – *b:* possibly a variable.
92. S Sgr:
   – possibly $x = m$.
94. R Sgr:
   – no information is available for the comparison stars.
95. V Sgr:
   – no information is available for the comparison stars.
96. R Sco:
   – $b = S$
   – *y:* one of the following stars:
      (1) $\alpha(2000) = 16^h17^m25^s.02$, $\delta(2000) = -22°49'07".08$, $B = 13^m.379$, and $V = 11^m.640$.
      (2) $\alpha(2000) = 16^h17^m30^s.64$, $\delta(2000) = -22°50'16".00$, $B = 13^m.420$, and $V = 12^m.810$.
97. S Sco
   – the same comparison stars as for R Sco.
98. R Set:
   – *d:* erroneous BD number: BD - 05 4836 should be replaced by BD - 05 4736;
   – *m:* erroneous BD number: BD - 05 4845 should be replaced by BD - 50 4745.
100. S Ser:
   – *m* and *n:* confused at the end of the observing period, possible erroneous nomenclature of the stars *m* and *n*.
101. T Ser:
   – *h':* erroneous BD number: BD + 06 3786 should be replaced by BD + 06 3776;
   – $a = e$;
   – $b = h$.
104. R Tau:
   – *l:* erroneous declination: +9°40'.5 should be replaced by +9°45'.5.
105. S Tau:
   – the star is absent on the map due to a possible brightness minimum.
106. T Tau:
   – *m:* the magnitude $V = 11^m.39$ was taken from the GCS catalog.
107. V Tau:
   – the coordinates of all comparison stars are unavailable.

109. S Uma:
— probable nomenclature confusion: *p, n',* and *k'*;
— *n':* the magnitude $V = 12^m.78$ was taken from the GCS catalog;
— *t:* the magnitude V = 12.17 was taken from the GCS catalog.

110. T Uma:
— the comparison stars are the same as for S UMa.

111. R Vir:
— *a:* erroneous BD number: BD + 8 2628 should be replaced by + 8 2626;
— *p:* erroneous declination: +7°54'7 should be replaced by + 8°04:7;
— the stars *g* and *g'* are confused with each other.

112. S Vir:
— *t, u,* and *v*: identified by the author using sketches in the text.

113. T Vir:
— *k:* erroneous declination: -05°11' should be replaced by -05°05';
— the stars *η* and *m* can be confused in the text;
— *m, n,* and *p:* identified by the author using the sketches in the text.

115. V Vir:
— *a* and *b:* a right-ascension error of up to 6 arcmin;
— *e* = W Vir: wrong BD number: BD - 02 3693 should be replaced by BD - 02 3683;
— *b*: is a binary, the total brightness: $B = 11^m.24$ and $V = 10^m.11$.

116. W Vir:
— the comparison stars are the same as for V Vir.

117. R Vul:
*h:* identified using a sketch in the text; a right-ascension error of 10 arcsec is possible;
*i, k, I, in, n:* the coordinates are absent in all sources; possible confusion at the end of the observing period since 1874;
*e:* possible 30-arcmin declination error: +22°51'. 1 should be replaced by +22°21'. 1.

The comprehensive list of identifications of comparison stars used by Schenfeld in observations of variables in the 19th century has been placed on the CrAO web site:
( http://www.craocrimea.ru/~aas/dbce.html ).

Photometric databases available at present give rise to the optimistic conviction that the light curves of various variable stars can be further refined based on new determinations of the magnitudes of comparison stars.

**Acknowledgments**
This research has made use of the SIMBAD database, operated at CDS, Strasbourg, France. The authors are also grateful to SAI staff members O. Bartunov, E. Rodichev, etc., for their celestial map, which is generally acknowledged as valuable, helpful research tool by many ex-USSR astronomers. We thank A.A. Shlyapnikov for helpful discussions during this work.